\documentclass[aps,prl,twocolumn,showpacs]{revtex4-1}
\usepackage{epsfig}
\usepackage{bm}

\newcommand{\be}{\begin{eqnarray}}
\newcommand{\ee}{\end{eqnarray}}
\newcommand{\ba}{\begin{array}}
\newcommand{\ea}{\end{array}}
\newcommand{\bmat}{\left(\begin{array}}
\newcommand{\emat}{\end{array}\right)}
\newcommand{\no}{\nonumber}

\newcommand{\diff}{\mathrm d}
\newcommand{\e}{\mathrm e}

\begin{document}
\title{From Classical Nonlinear Integrable Systems 
to Quantum Shortcuts to Adiabaticity}
\author{Manaka Okuyama}
\author{Kazutaka Takahashi}
\affiliation{Department of Physics, Tokyo Institute of Technology, 
Tokyo 152-8551, Japan}

\date{\today}

\begin{abstract} 
Using shortcuts to adiabaticity, we solve 
the time-dependent Schr\"odinger equation 
that is reduced to a classical nonlinear integrable equation.
For a given time-dependent Hamiltonian, 
the counterdiabatic term is introduced 
to prevent nonadiabatic transitions.
Using the fact that the equation for the dynamical invariant 
is equivalent to the Lax equation in nonlinear integrable systems, 
we obtain the counterdiabatic term exactly.
The counterdiabatic term is available 
when the corresponding Lax pair exists 
and the solvable systems are classified 
in a unified and systematic way.
Multisoliton potentials obtained from the Korteweg--de Vries equation and 
isotropic $XY$ spin chains from the Toda equations are studied in detail.
\end{abstract}
\maketitle

{\it Introduction.}--
Ideal control of quantum systems 
has attracted interest recently from both theoretical and 
practical viewpoints.
Rapid technological advances make it possible to manipulate 
quantum systems precisely, and 
designing the optimal Hamiltonian is a realistic important problem.
The meaning of optimality is not obvious, 
and various methods have been proposed theoretically
in different contexts.
In the methods using shortcuts to adiabaticity, 
the Hamiltonian is designed so that the state follows 
an adiabatic passage of 
a reference Hamiltonian~\cite{DR1, DR2, Berry, CRSCGM, STA}.
This technique was realized 
in several experiments~\cite{SSVL, SSCVL, Betal, Zetal}
and is expected to be applied to 
the adiabatic quantum computation called quantum annealing~\cite{KN}.

Although the formulation of the method is general, 
explicit constructions of the Hamiltonian
are restricted to simple systems 
such as two- and three-level systems~\cite{Berry, CLRGM}, 
harmonic oscillators~\cite{MCILR},
and scale-invariant systems~\cite{delCampo}.
Most of systems fall into these categories, 
and various applications have been studied in many 
works~\cite{STA, delCampo2, BCSM, JTMMP, KPRO, PGAHM, PS}.
The question addressed in this Letter is 
whether there are other classes of Hamiltonians  
in which the exact solutions are available.
We propose a new method 
using the results from nonlinear integrable systems 
such as Korteweg--de Vries (KdV) solitons~\cite{KdV} and 
Toda lattices~\cite{Toda}.
In nonlinear systems, a quantum mechanical viewpoint is used  
to solve the equations of motion.
We find that the method developed there 
can be directly applied to the quantum adiabatic dynamics.
Several previous works used the result from the integrable systems
to construct a driver Hamiltonian~\cite{TGRM, Masuda, delCampo}.
Here we utilize the integrability of classical nonlinear systems 
to develop the fundamental principle 
to design the quantum Hamiltonian in a systematic way.
It is applied not only to a specific system 
but also to a broad class of integrable systems
including many-body systems. 

{\it Counterdiabatic driving and dynamical invariant.}--
In counterdiabatic driving, 
known as one of the strategies in shortcuts to adiabaticity, 
we control the adiabatic states of 
the time-dependent Hamiltonian $\hat{H}_{\rm ad}(t)$.
The adiabatic state is defined by a linear combination 
of the instantaneous eigenstates $|n(t)\rangle$ of $\hat{H}_{\rm ad}(t)$ 
as $|\psi(t)\rangle=\sum_n c_n\e^{-i\theta_n(t)}|n(t)\rangle$,
where $c_n$ is a time-independent constant and 
$\theta_n(t)=\int \diff t\,\left(\langle n(t)|\hat{H}_{\rm ad}(t)|n(t)\rangle
-i\langle n(t)|\frac{\partial}{\partial t}|n(t)\rangle\right)$ 
represents a time-dependent phase~\cite{ad}.
This state does not satisfy the Schr\"odinger equation 
owing to nonadiabatic transitions.
Those unwanted transitions are suppressed 
by introducing an additional term called the 
counterdiabatic term, $\hat{H}_{\rm cd}(t)$.
The adiabatic state of $\hat{H}_{\rm ad}(t)$ 
satisfies the Schr\"odinger equation: 
\be
 i\frac{\partial}{\partial t}|\psi(t)\rangle=\left(\hat{H}_{\rm ad}(t)
 +\hat{H}_{\rm cd}(t)\right)|\psi(t)\rangle.
\ee
The spectral representation of $\hat{H}_{\rm cd}(t)$ is 
given by~\cite{DR1, DR2, Berry}
\be
 \hat{H}_{\rm cd}(t)=i\sum_n 
 \left(1-|n(t)\rangle\langle n(t)|\right)
 \frac{\partial|n(t)\rangle}{\partial t}\langle n(t)|. \label{cd}
\ee

Various strategies exist for designing the Hamiltonian 
in shortcuts to adiabaticity.
The common property is described by the existence of 
the Lewis-Riesenfeld dynamical invariant $\hat{F}(t)$ satisfying 
\be
 i\frac{\partial\hat{F}(t)}{\partial t}
 =[\hat{H}(t),\hat{F}(t)], \label{di}
\ee
where $\hat{H}(t)$ is the total Hamiltonian.
It was shown that 
the eigenvalues of $\hat{F}(t)$ are independent of $t$, 
and the solutions of the Schr\"odinger equation are given by adiabatic
states of $\hat{F}(t)$~\cite{LR}.
This means that the Hamiltonian is divided into adiabatic 
[$\hat{H}_{\rm ad}(t)$] and counterdiabatic [$\hat{H}_{\rm cd}(t)$] components, 
and $\hat{F}(t)$ commutes with $\hat{H}_{\rm ad}(t)$.
$\hat{H}_{\rm cd}(t)$ is written using the eigenstates 
of $\hat{F}(t)$ as Eq.~(\ref{cd}). 
In inverse engineering~\cite{CRSCGM}, 
the problem is formulated using Eq.~(\ref{di}) 
instead of Eq.~(\ref{cd}), as we show below.
Furthermore, 
the optimality of the counterdiabatic driving is shown 
using a method called the quantum brachistochrone~\cite{CHKO, Takahashi1}.
The optimal solution is represented by the invariant, 
which is constructed for given constraints of the Hamiltonian.
Thus, finding the invariant is the key to designing the optimal Hamiltonian.

We mainly treat a class of systems such that
$\hat{F}(t)=\hat{H}_{\rm ad}(t)$.
The counterdiabatic term $\hat{H}_{\rm cd}(t)=\hat{H}(t)-\hat{H}_{\rm ad}(t)$
is obtained by solving Eq.~(\ref{di}). 
In this case, the eigenvalues of $\hat{H}_{\rm ad}(t)$ are time independent.
$\hat{H}_{\rm cd}(t)$ is expressed by a linear combination 
of possible operators, and the time dependence of the coefficients is 
obtained by solving Eq.~(\ref{di}).
The series is generally infinite, 
and solving the equation is a formidable task.
However, in the integrable systems that we discuss in this Letter, 
the counterdiabatic term is expressed in a compact form.

{\it Lax form for KdV equation.}--
First, we study one-dimensional systems 
in a time-dependent potential $u(x,t)$.
The adiabatic Hamiltonian is given by
\be
 \hat{H}_{\rm ad}(t)=\hat{p}^2+u(\hat{x},t). \label{Had}
\ee
The potential is left undetermined, and we study conditions 
that a compact form of the counterdiabatic term is obtained.
As the simplest case, if we assume that $\hat{H}_{\rm cd}(t)$ includes terms 
up to first order in $\hat{p}$, the general form is given by 
$\hat{H}_{\rm cd}(t) = v(t)\hat{p}+\epsilon(t)$, where 
$v(t)$ and $\epsilon(t)$ are arbitrary functions.
The potential must have the form 
$u(x,t)=u_0(x-x_0(t))$, where $u_0$ is an arbitrary function with a single 
variable, and $x_0(t)=\int v(t)\diff t$.
A similar result is obtained for $\hat{H}_{\rm cd}(t)$
including second-order terms in $\hat{p}$~\cite{sm-kdv}.
These results are well known and 
are described by the method of scale-invariant driving~\cite{delCampo}.

A novel solution is obtained when we consider the third-order term.
The counterdiabatic term takes the form 
\be
 \hat{H}_{\rm cd}(t)=a(t)\left[\hat{p}^3
 +\frac{3}{4}\left(\hat{p}u(\hat{x},t)+u(\hat{x},t)\hat{p}\right)
 \right]+c_1(t)\hat{p},  \no\\\label{Hcd}
\ee
where $a(t)$ and $c_1(t)$ are arbitrary time-dependent functions.
We neglected trivial contributions that are proportional to 
$\hat{H}_{\rm ad}(t)$ and the identity operator.
The condition for the potential $u(x,t)$ reads 
\be
 \frac{\partial u}{\partial t} = -\frac{a(t)}{4}\left(
 6u\frac{\partial u}{\partial x}
 -\frac{\partial^3 u}{\partial x^3}\right)
 -2c_1(t)\frac{\partial u}{\partial x}. 
\ee
Using the change of variables, 
we set $a(t)=-4$ and $c_1(t)=0$ in the following analysis~\cite{sm-kdv}.
Then, this equation is reduced to the standard form of 
the KdV equation~\cite{KdV}.
The KdV equation is known as a typical nonlinear integrable system
and has multisoliton solutions~\cite{GGKM}.

It is no accident that the KdV equation is obtained
in the present formulation.
The integrability of nonlinear systems is generally 
represented by the Lax equation, 
\be
 \frac{\partial L(t)}{\partial t}=[M(t),L(t)],
\ee
where the set of operators $(L(t),M(t))$ is called the Lax pair~\cite{Lax}.
The Lax equation is equivalent to the equation 
for the dynamical invariant Eq.~(\ref{di}).
The eigenvalues of $L(t)$ are independent of $t$.
Then, $M(t)$ is defined as  
$\partial_t\psi(t)=M(t)\psi(t)$,
where $\psi(t)$ is an eigenfunction of $L(t)$. 
The Lax pair for the KdV equation is known to have the form 
\be
 && L(t)=-\frac{\partial^2}{\partial x^2}+u(x,t), \no\\
 && M(t)= -4\frac{\partial^3}{\partial x^3}+3\frac{\partial}{\partial x}u(x,t)
 +3u(x,t)\frac{\partial}{\partial x}.
\ee
As we see from the comparison between the Lax form and 
the dynamical invariant,
$L(t)$ and $M(t)$ correspond to    
$\hat{H}_{\rm ad}(t)$ in Eq.~(\ref{Had}) and 
$-i\hat{H}_{\rm cd}(t)$ in Eq.~(\ref{Hcd}), respectively.
In integrable systems, there exist infinite conserved quantities,
which reflect the property that
the eigenvalues of the dynamical invariant are independent of $t$.

In the same way, considering higher-order terms, 
we can find a hierarchical structure in the KdV equations.
It is generally known that a different KdV equation is obtained 
at each odd order~\cite{Lax}.
For example, at fifth order, 
the counterdiabatic term is given by 
$H_{\rm cd}(t)=-16p^5+20(p^3u+up^3)-30upu-5(pu_{xx}+u_{xx}p)$,  
where the subscript denotes differentiation in terms of the variable, 
and the potential satisfies 
$u_t-10(u_{xxx}u+2u_{xx}u_x)+30u^2u_x+u_{xxxxx}=0$.

{\it Deformation of the counterdiabatic term.}--
Although we have obtained the exact form of 
the counterdiabatic term Eq.~(\ref{Hcd}), 
it contains a cubic term in $\hat{p}$ and is not convenient for 
our purpose of quantum control.
This drawback is circumvented by using gauge transformation and
state-dependent deformation~\cite{ICTMR, delCampo, MTCM, Takahashi2}.
We demonstrate this by treating the single- and double-soliton potentials.

The single-soliton solution of the KdV equation is given by 
$u(x,t) = -2\kappa^2/\cosh(\kappa x-4\kappa^3 t)$,
where $\kappa$ is a real positive parameter.
The adiabatic Hamiltonian [Eq.~(\ref{Had})] with this hyperbolic Scarf potential 
has a single bound state whose eigenenergy is given by $E_0=-\kappa^2$.
To deform the counterdiabatic term, we rewrite Eq.~(\ref{Hcd}) 
with $a(t)=-4$ and $c_1(t)=0$ as 
\be
 \hat{H}_{\rm cd}(t) &=& -4(\hat{p}+iW(\hat{x},t))\hat{p}
 (\hat{p}-iW(\hat{x},t)) \no\\
 && -\left(\hat{p}\tilde{u}(\hat{x},t)+\tilde{u}(\hat{x},t)\hat{p}\right)
 -4E_0\hat{p},  \label{Hcd2}
\ee
where $W(x,t)=\kappa\tanh(\kappa x-4\kappa^3 t)$ 
and $\tilde{u}(x,t)=W^2(x,t)+\partial_xW(x,t)+E_0$.
The adiabatic Hamiltonian in this case has 
supersymmetry~\cite{Witten, ABM, GSS, CKS, sm-susy}, and 
the potential is written using the superpotential $W(x,t)$ as
$u(x,t)=W^2(x,t)-\partial_xW(x,t)+E_0$.
Supersymmetry also tells us that 
the ground-state eigenfunction $\psi(x,t)$
satisfies the differential equation $(-i\partial_x-iW(x,t))\psi(x,t)=0$.
This means that the first term in Eq.~(\ref{Hcd2}) is neglected 
if we treat only the ground state.
Furthermore, the superpartner potential $\tilde{u}(x,t)$
goes to zero in this case.
Thus, we arrive at a simple form of the counterdiabatic term: 
$\hat{H}_{\rm cd}(t)=-4E_0\hat{p}=4\kappa^2 \hat{p}$, 
which only affects the phase of the wave function 
and does not change the probability density.
This form can be derived from the formula 
for scale-invariant systems~\cite{delCampo}.
In the single-soliton potential, the system has translational symmetry,
and we can directly obtain this simple form.

For multisoliton potentials, the system is not scale invariant, 
and our method becomes useful.
The double-soliton potential is represented as 
\be
 u(x,t)&=&-2\frac{\partial^2}{\partial x^2}\ln
 \biggl[1+A_1\e^{2\eta_1}+A_2\e^{2\eta_2}
 \no\\
 && +\left(\frac{\kappa_1-\kappa_2}{\kappa_1+\kappa_2}\right)^2
 A_1A_2\e^{2(\eta_1+\eta_2)}\biggr],
\ee
where $\kappa_1$, $\kappa_2$, $A_1$, and $A_2$ 
are real positive parameters,
and $\eta_{i}=\kappa_ix-4\kappa_i^3t$ with $i=1,2$.
This system has two bound states.
When $\kappa_1>\kappa_2$, the ground-state energy is $E_0=-\kappa_1^2$ 
and the first excited-state energy is $E_1=-\kappa_2^2$.
The ground state is localized in the deeper well,
and the first excited state is in the other well. 

It is also possible to consider the deformation of 
the counterdiabatic term [Eq.~(\ref{Hcd})]  
using the representation Eq.~(\ref{Hcd2})~\cite{sm-deform}.
When we consider the ground state, we can again neglect 
the first term of Eq.~(\ref{Hcd2}).
Using a gauge transformation, we obtain the counterdiabatic potential 
\be
 && V_{\rm cd}(x,t) = -\tilde{u}^2(x,t)
 +4(\kappa_1^2-\kappa_2^2)\tilde{u}(x,t), 
\ee
where $\tilde{u}(x,t)=-2\kappa_2^2/\cosh^2(\eta_2+\delta_2)$, with 
$\delta_2=\ln\left[A_2(\kappa_1-\kappa_2)/(\kappa_1+\kappa_2)\right]/2$.
Deformation of the excited states can be obtained in a similar way.

\begin{center}
\begin{figure}[t]
\begin{center}
\includegraphics[width=0.8\columnwidth]{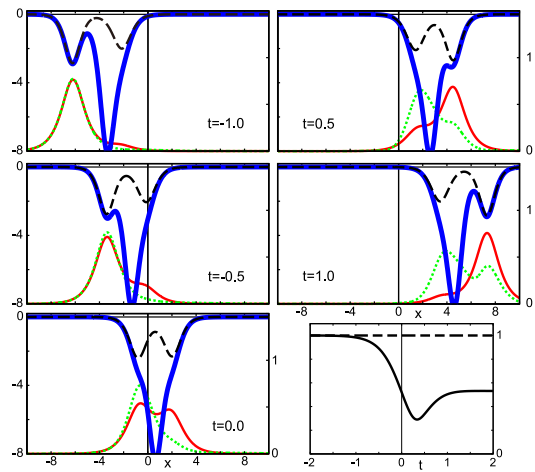}
\end{center}
\caption{Time evolution of double-soliton potential.
Plotted are the soliton potential $u(x,t)$ (black dashed line, 
left scale), 
total potential $u(x,t)+V_{\rm cd}(x,t)$ (blue thick line, left), and 
ground-state wave function $|\psi(x,t)|$ (red solid line, right).
Green dotted lines represent the wave function $|\psi_0(x,t)|$
without the counterdiabatic potential $V_{\rm cd}(x,t)$ 
calculated numerically (right-hand scale).
Solid line in the lower right-hand panel represents the overlap 
between states with and without the counterdiabatic term 
$|\langle\psi_0(t)|\psi(t)\rangle|$.
We take $A_1=A_2=3.0$, $\kappa_1=1.2$, and $\kappa_2=1.0$.
}
\label{fig1}
\end{figure}
\end{center}

We plot the potential and ground-state wave function in Fig.~\ref{fig1}.
The counterdiabatic potential is represented by the single soliton 
$\tilde{u}(x,t)$ with the argument $\eta_2+\delta_2$.
It makes a deep well around the shallower well
in the original potential 
and moves so that the moving state, trapped in the deep well 
of the original potential, 
is not disturbed by the shallow well. 

The above demonstrations show that 
we can consider particle transport by using the soliton potentials.
A particle is conveyed by a soliton potential, and 
we can find conditions such that 
the state is not disturbed by noises represented by other solitons.
The use of the soliton potential shows 
that the control of systems is possible by changing the potential locally,
which can be a useful manipulating method.
The exact solutions from the KdV equation also have a great advantage that 
the small deviations can be treated perturbatively
to study stability~\cite{KM}.
This method can be a principle to design the particle transport 
in general complicated systems.

{\it Free fermion, spin chain, and Toda lattice.}--
Next, we demonstrate the design of the Hamiltonian 
by using a result from the Lax form in nonlinear integrable systems.
A famous example is the Toda lattice system.
This one-dimensional classical lattice system 
has nearest-neighbor hopping in an exponential form, and 
the Lax form is written in a matrix form~\cite{Toda, Flaschka, sm-toda}.
The matrix has a tridiagonalized form and
can be interpreted as the quantum Hamiltonian 
for a one-dimensional lattice system with nearest-neighbor hopping: 
\be
 \hat{H}_{\rm ad}(t) &=& \sum_{n=1}^N J_n(t)\left(
 \hat{f}_n^\dag \hat{f}_{n+1}^{\phantom{\dag}}
 +\hat{f}_{n+1}^\dag \hat{f}_{n}^{\phantom{\dag}}\right) \no\\ &&
 +\sum_{n=1}^N h_n(t)\left(\hat{f}_n^\dag \hat{f}_n^{\phantom{\dag}}
 -\frac{1}{2}\right), \no\\
 \hat{H}_{\rm cd}(t) &=& i\sum_{n=1}^N J_n(t)\left(
 \hat{f}_n^\dag \hat{f}_{n+1}^{\phantom{\dag}}
 -\hat{f}_{n+1}^\dag \hat{f}_{n}^{\phantom{\dag}}\right). \label{fermion}
\ee
The fermion operators $\hat{f}_n$ and $\hat{f}_n^\dag$ 
satisfy anticommutation relations~\cite{sm-jw}.
Substituting these expressions into Eq.~(\ref{di}) 
with $\hat{F}(t)=\hat{H}_{\rm ad}(t)$, 
we find the Toda equations
\be
 && \frac{\diff J_n(t)}{\diff t}=J_n(t)(h_{n+1}(t)-h_n(t)), \no\\
 && \frac{\diff h_n(t)}{\diff t}=2(J_n^2(t)-J_{n-1}^2(t)). \label{Todaeqs}
\ee
The solutions of the Toda equations are discussed below.

Although this transformation is a straightforward task, 
the resulting Hamiltonian has the time-dependent hopping amplitude and 
is generally difficult to realize.
The matrix representation of the Lax form reminds us that 
the Hamiltonian is interpreted as a quantum spin system.
By using the Jordan-Wigner transformation~\cite{JW, LSM}, 
we can map the free fermion onto the $XY$ spin model~\cite{sm-jw}:
\be
 && \hat{H}_{\rm ad}(t)=\sum_{n=1}^N\frac{J_n(t)}{2}\left(
 \hat{\sigma}^x_n\hat{\sigma}^x_{n+1}
 +\hat{\sigma}^y_n\hat{\sigma}^y_{n+1}\right)
 +\sum_{n=1}^N\frac{h_n(t)}{2}\hat{\sigma}_n^z, \no\\
 &&\hat{H}_{\rm cd}(t)=\sum_{n=1}^N\frac{J_n(t)}{2}\left(
 \hat{\sigma}_n^x\hat{\sigma}_{n+1}^y
 -\hat{\sigma}_n^y\hat{\sigma}_{n+1}^x\right), \label{hxy}
\ee
where $\{\hat{{\bm \sigma}}_n
=(\hat{\sigma}_n^x,\hat{\sigma}_n^y,\hat{\sigma}_n^z)\}$ 
represent the Pauli matrices at each site $n$.

It is instructive to see that the isotropic, nonuniform interaction 
$J_n(t)$ is necessary to obtain the adiabatic control.
For a uniform system with site-independent couplings, 
the energy eigenvalues and eigenstates 
of the $XY$ spin Hamiltonian $\hat{H}_{\rm ad}(t)$ are obtained exactly 
by using the Jordan-Wigner transformation~\cite{dCZ, Takahashi3, SOMdC, Damski}.
In that case, 
we can find $\hat{H}_{\rm cd}(t)$ by using Eq.~(\ref{cd}).
The problem is that $\hat{H}_{\rm cd}(t)$ has 
a noncompact form including many-body interaction terms~\cite{dCZ, Takahashi3}.
We circumvent this difficulty by using the integrability of the Toda lattice.

It is interesting to compare the Lax pair for the spin model
with that for the Toda lattice.
Each term in Eq. (\ref{hxy}) commutes with 
the total magnetization in the $z$ direction: $\hat{M}=\sum_n\hat{\sigma}_n^z$.
In the single spin-flip sectors with $M=\pm(N-2)$, 
the matrix representation of the Hamiltonian is 
the same as that of the Lax pair for the Toda lattice.
There are no counterparts for other sectors, and 
it is possible to find new Lax pairs for classical nonlinear systems 
by analyzing those matrix representations.

We study several typical solutions of the Toda equations (\ref{Todaeqs})
to see the behavior of the states in the corresponding spin system.
For $N=3$ with the open boundary condition, 
one of the solutions is given as follows.
$J_1(t)$ and $h_1(t)$ are given by 
\be
 && J_1(t)=vv_1\frac{\sqrt{v_1^2+v_2^2\cosh(2vt)}}
 {v_1^2\cosh(2vt)+v_2^2}, \no\\
 && h_1(t) = v\frac{v_1^2\sinh(2vt)}{v_1^2\cosh(2vt)+v_2^2},
\ee
where $v_1$ and $v_2$ are constants and $v=\sqrt{v_1^2+v_2^2}$.
$J_2(t)$ is obtained by interchanging $v_1$ and $v_2$ in $J_1(t)$, 
$h_3(t)$ is obtained by interchanging $v_1$ and $v_2$ in $-h_1(t)$,
and $h_2(t) = -h_1(t)-h_3(t)$.
We note that $(J_1(0),J_2(0))=(v_1, v_2)$,  
$(h_1(0), h_2(0), h_3(0))=(0, 0, 0)$, 
$(J_1(\infty),J_2(\infty))=(0, 0)$, and  
$(h_1(\infty), h_2(\infty), h_3(\infty))=(v, 0, -v)$.
These limiting values show that the Hamiltonian of the $XY$ model 
can be continuously deformed to the noninteracting Hamiltonian
with the same eigenvalues by using the time evolution of the Toda equations.
This result is generalized to an arbitrary number of spins $N$~\cite{Moser}.
We note that continuous change is possible 
for the isotropic $XY$ model and 
not for anisotropic models, including the Ising model,
which is considered to be 
related to the universal adiabatic computing property of 
the $XY$ Hamiltonian~\cite{BL}.

\begin{center}
\begin{figure}[t]
\begin{center}
\includegraphics[width=0.8\columnwidth]{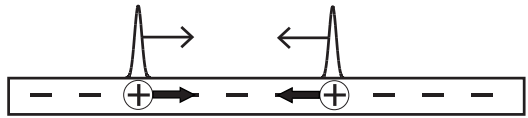}
\end{center}
\caption{Transport of flipped spin by soliton interactions.
$+$ or $-$ represents spin state at each site and 
the soliton wave represents the coupling functions.
The figure represents the transport of two flipped spins 
by double solitons and 
other cases can be described in a similar way~\cite{sm-xy}.
}
\label{fig2}
\end{figure}
\end{center}

In infinite systems, the boundary effect is neglected and
the soliton solutions are obtained 
by using the inverse scattering method~\cite{Flaschka}.
The solutions allow us to control the spin state.
As we find in the KdV solitons, the flipped spin is transported 
by controlling the spin interaction and the magnetic field
as in Fig.~\ref{fig2}~\cite{sm-xy}.
The quantum state transfer has been discussed in many contexts 
and the use of spin chain is one of the typical applications~\cite{Bose}. 
Our result based on the soliton dynamics 
can be useful as a robust control method.

For practical applications, one of the disadvantages of 
counterdiabatic driving is 
that we need to introduce an additional term $\hat{H}_{\rm cd}(t)$.
It is possible to perform the gauge transformation
to keep the total Hamiltonian in a simple form.
This change of representation is interpreted as inverse engineering.
The concept of shortcuts to adiabaticity extends beyond 
counterdiabatic driving, 
and we can consider other formulations.
In inverse engineering, the Hamiltonian is 
kept in the original form and the invariant is constructed
from Eq.~(\ref{di})~\cite{sm-inv}.

{\it Nonisospectral Hamiltonian.}-- 
Although our discussion is restricted to systems with 
isospectral Hamiltonian, 
it can be generalized to nonisospectral systems.
The isospectral Hamiltonian is obtained
by assuming $\hat{F}(t)=\hat{H}_{\rm ad}(t)$.
In classical nonlinear integrable systems, 
the Lax pair is generalized to nonisospectral systems~\cite{ZY}.
The integrability is maintained when we consider the case 
$\hat{F}(t)= \gamma^2(t)\hat{H}_{\rm ad}(t)$, where 
$\gamma(t)$ is a time-dependent function.
In systems with the Hamiltonian Eq.~(\ref{Had}), 
if we impose this condition, we can find a generalized KdV hierarchy.
In the simplest ansatz where the counterdiabatic term 
is linear in $\hat{p}$, we can find the potential 
form $u(x,t)=\gamma^{-2}(t)u_0((x-x_0(t))/\gamma(t))$
with the counterdiabatic term 
$\hat{H}_{\rm cd}(t)\propto \hat{x}\hat{p}+\hat{p}\hat{x}$.
This is known as the scale-invariant driving~\cite{delCampo}.
The analysis can be applied to higher-order hierarchies 
to give results that cannot be described 
by the scale-invariant driving~\cite{sm-noniso}.
Thus, the previous known results can be described in the present formulation
in a systematic way and further generalizations are possible.

{\it Conclusions.}--
We have proposed counterdiabatic driving 
for nonlinear integrable systems.
The Lax form is known for various systems 
other than the KdV and Toda systems.
Possible applications are the modified KdV equation, 
sine-Gordon equation, 
nonlinear Schr\"odinger equation, 
Ablowitz-Kaup-Newell-Segur formalism~\cite{AKNS}, and so on.
We stress that the infinite set of Lax pairs 
is known in nonlinear integrable systems, and 
one can find a corresponding counterdiabatic driving in each system.

Our approach establishes a fruitful connection between 
the classical nonlinear systems and the dynamical quantum systems.
In the quantum systems, 
the counterdiabatic term was available only for limited simple systems.
In the present work, the solvable systems can be obtained 
in a unified and systematic way
by using the knowledge from the classical nonlinear integrable systems.
For example, in the potential systems 
where $\hat{H}_{\rm ad}(t)$ is written as Eq.~(\ref{Had}),
the counterdiabatic term is obtained from the KdV hierarchy.
In the example of the Toda systems, 
we can find the corresponding quantum spin system and fermion system 
by knowing the matrix representation of the Lax form.
The generalization is possible and 
we can use results from systems such as 
the Calogero-Sutherland and Calogero-Moser models 
and Volterra lattice~\cite{Sutherland, Calogero, Moser2, KvM}.
Several applications of such systems will be reported elsewhere.

The authors are grateful to Katsuhiro Nakamura and Hidetoshi Nishimori 
for useful discussions.
We acknowledge financial support from the 
ImPACT Program of the Council for Science, 
Technology, and Innovation, Cabinet Office, Government of Japan.
K.T. was supported by JSPS KAKENHI Grant No. 26400385.


\onecolumngrid

\def\theequation{S\arabic{equation}}
\makeatletter
\@addtoreset{equation}{section}
\makeatother

\setcounter{equation}{0}

\newpage
\section{Supplemental Material}

\twocolumngrid
\section{KdV equation}

For the adiabatic Hamiltonian 
\be
 \hat{H}_{\rm ad}(t)=\hat{p}^2+u(\hat{x},t), \label{Had-s}
\ee
where $u$ is a potential function determined below, 
we seek the counterdiabatic term $\hat{H}_{\rm cd}(t)$ 
satisfying the equation for the invariant: 
\be
 i\frac{\partial\hat{H}_{\rm ad}(t)}{\partial t}
 =[\hat{H}_{\rm cd}(t),\hat{H}_{\rm ad}(t)]. \label{di2}
\ee
As a first ansatz, we write the form 
\be
 \hat{H}_{\rm cd}(t)=v(t)\hat{p}+\epsilon(\hat{x},t).
\ee
Then, we obtain from Eq.~(\ref{di2})
\be
 && \frac{\partial\epsilon(x,t)}{\partial x}=0, \\
 && \frac{\partial u(x,t)}{\partial t}
 +v(t)\frac{\partial u(x,t)}{\partial x}=0.
\ee
The first equation shows that $\epsilon(x,t)$ is independent of $x$.
From the second equation, 
the functional form of the potential is obtained as  
\be
 u(x,t)=u_0(x-x_0(t)), \qquad \frac{\diff x_0(t)}{\diff t}=v(t).
\ee
This potential describes time-dependent translational motions.
In this case, the counterdiabatic term is obtained as  
\be
 \hat{H}_{\rm cd}(t)=v(t)\hat{p}+\epsilon(t), \label{firstp}
\ee
where $\epsilon(t)$ is an arbitrary function.
This result is known as scale-invariant driving.
We see from Eq.~(\ref{di2}) that $\epsilon(t)$ is 
always arbitrary in the present formulation.

Next, we study $\hat{H}_{\rm cd}(t)$ including terms up to second 
order in $\hat{p}$.
We write 
\be
 \hat{H}_{\rm cd}(t)=a(t)\hat{p}^2
 +\hat{p}b(\hat{x},t)+b(\hat{x},t)\hat{p}+\epsilon(\hat{x},t).
\ee
A calculation similar to the previous one gives us 
$\hat{H}_{\rm cd}(t)$ in the form 
\be
 \hat{H}_{\rm cd}(t) 
 = v(t)\hat{p}+\epsilon(t)+a(t)\left(\hat{p}^2+u(\hat{x},t)\right),
\ee
and the potential has the scale-invariant form 
\be
 u(x,t)=u_0(x-x_0(t)), \qquad \frac{\diff x_0(t)}{\diff t}=v(t).
\ee
The first term of $\hat{H}_{\rm cd}(t)$ 
can be obtained from the formula for scale-invariant driving,
and the last term is proportional to the adiabatic Hamiltonian (\ref{Had-s}).
Thus, in the second-order case, we do not have any interesting result.

A nontrivial result is obtained when we consider third-order terms.
We obtain the most general form 
\be
 && \hat{H}_{\rm cd}(t) 
 = \frac{a(t)}{4}\left[
 4\hat{p}^3+3\left(\hat{p}u(\hat{x},t)+u(\hat{x},t)\hat{p}\right)
 \right] \no\\
 && +b(t)\left(\hat{p}^2+u(\hat{x},t)\right)
 +c_1(t)\hat{p}+c_0(t). \label{cubicp}
\ee
The second and last terms make trivial contributions and are neglected.
The potential function must satisfy 
\be
 \frac{\partial u}{\partial t} = -\frac{a(t)}{4}\left(
 6u\frac{\partial u}{\partial x}
 -\frac{\partial^3 u}{\partial x^3}\right)
 -c_1(t)\frac{\partial u}{\partial x}. 
 \label{kdvc}
\ee
This equation is reduced to the KdV equation by setting 
\be
 u(x,t)=\tilde{u}(x-x_0(t),t), \quad \frac{\diff x_0(t)}{\diff t}=c_1(t).
\ee
Here, $\tilde{u}$ satisfies 
\be
 \frac{\partial \tilde{u}}{\partial t} = -\frac{a(t)}{4}\left(
 6\tilde{u}\frac{\partial \tilde{u}}{\partial x}
 -\frac{\partial^3 \tilde{u}}{\partial x^3}\right).
\ee
By setting $a(t)=-4$, we obtain the standard form of the KdV equation.
\be
 \frac{\partial \tilde{u}}{\partial t} = 
 6\tilde{u}\frac{\partial \tilde{u}}{\partial x}
 -\frac{\partial^3 \tilde{u}}{\partial x^3}. \label{kdv-s}
\ee

\section{Supersymmetry}

In this section, we summarize supersymmetry and its consequences
in quantum mechanics.
We consider the Hamiltonian 
\be
 \hat{H}^{(+)} 
 &=& (\hat{p}+iW(\hat{x}))(\hat{p}-iW(\hat{x})) \no\\
 &=& \hat{p}^2+W^2(\hat{x})-W'(\hat{x}),
\ee
where the real function $W(x)$ is called the superpotential, and 
$W'(x)$ denotes the derivative of $W(x)$.
This Hamiltonian is in a factorized form and 
has nonnegative energy eigenvalues because of the relation 
\be
 \hat{H}^{(+)} = (\hat{p}-iW(\hat{x}))^\dag(\hat{p}-iW(\hat{x})).
\ee
The eigenvalue equation is written as 
\be
 \hat{H}^{(+)}|n^{(+)}\rangle = E_n|n^{(+)}\rangle.
\ee

Then, we define the superpartner Hamiltonian as 
\be
 \hat{H}^{(-)} 
 &=& (\hat{p}-iW(\hat{x}))(\hat{p}+iW(\hat{x})) \no\\
 &=& \hat{p}^2+W^2(\hat{x})+W'(\hat{x}).
\ee
It is straightforward to show that this Hamiltonian 
has the energy eigenstate 
\be
 |n^{(-)}\rangle = \frac{1}{\sqrt{E_n}}
 \left(\hat{p}-iW(\hat{x},t)\right)|n^{(+)}\rangle \label{npm}
\ee
and the energy eigenvalue $E_n$.
This means that $\hat{H}^{(+)}$ and $\hat{H}^{(-)}$ have 
the same energy spectrum for positive-energy states.
It is also possible to write 
\be
 |n^{(+)}\rangle = \frac{1}{\sqrt{E_n}}
 \left(\hat{p}+iW(\hat{x},t)\right)|n^{(-)}\rangle. \label{nmp}
\ee

We note that Eqs.~(\ref{npm}) and (\ref{nmp}) cannot be used 
for the zero-energy state,
which is defined as 
\be
 \left(\hat{p}-iW(\hat{x},t)\right)|0^{(+)}\rangle = 0. \label{zero}
\ee
The corresponding wavefunction is represented as 
\be
 \psi_0^{(+)}(x) = C\exp\left(-\int W(x)\diff x\right). \label{gs}
\ee
If this function is normalized, the zero-energy state exists.
In that case, the superpartner cannot be defined, and 
we conclude that the zero-energy state breaks supersymmetry.

As an example, we consider the superpotential
\be
 W(x)=\kappa\tanh (\kappa x).
\ee
Then, we obtain
\be
 && \hat{H}^{(+)}=\hat{p}^2
 -\frac{2\kappa^2}{\cosh^2(\kappa \hat{x})}+\kappa^2, \\
 && \hat{H}^{(-)}=\hat{p}^2+\kappa^2.
\ee
The potential in $\hat{H}^{(+)}$ represents a single soliton
derived from the KdV equation (\ref{kdv-s}).
This example shows that the Hamiltonian with the single-soliton potential 
and the free-particle Hamiltonian form a superpartner pair.
$\hat{H}^{(+)}$ has a single bound state with zero energy.
The scattering states of $\hat{H}^{(+)}$ are transformed to 
the free-particle scattering states of $\hat{H}^{(-)}$ by Eq.~(\ref{npm}).
The ground state wavefunction of $\hat{H}^{(+)}$ is given by 
\be
 \psi_0^{(+)}(x) = \frac{\sqrt{\frac{\kappa}{2}}}{\cosh (\kappa x+\delta)},
\ee
where $\delta$ is an arbitrary constant.

In the next section, 
we show the solution of the double-soliton potential as a nontrivial example.

\section{Deformation of the counterdiabatic term in a double-soliton system}

In this section, we study the deformation of 
the double-soliton potential
\be
 && u(x,t) = -2\frac{\partial^2}{\partial x^2}\ln
 \biggl[1+A_1\e^{2\eta_1}+A_2\e^{2\eta_2} \no\\
 && +\left(\frac{\kappa_1-\kappa_2}{\kappa_1+\kappa_2}\right)^2
 A_1A_2\e^{2(\eta_1+\eta_2)}\biggr], \label{double}
\ee
where $\kappa_1$, $\kappa_2$, $A_1$, and $A_2$ 
are real positive parameters, 
and $\eta_{i}=\kappa_ix-4\kappa_i^3t$ with $i=1,2$.
As we mentioned in the main text, 
this potential satisfies the KdV equation and
the system has two bound states.
When $\kappa_1>\kappa_2$, the ground-state energy is $E_0=-\kappa_1^2$,
and the first excited-state energy is $E_1=-\kappa_2^2$.

This system also has supersymmetry, and 
the superpotential is written as 
\be
 && W(x,t)=-\kappa_1 
 +\frac{\partial}{\partial x} \no\\
 &&\times\ln\biggl[
 \frac{1+A_1\e^{2\eta_1}+A_2\e^{2\eta_2}
 +\left(\frac{\kappa_1-\kappa_2}{\kappa_1+\kappa_2}\right)^2
 A_1A_2\e^{2(\eta_1+\eta_2)}}{1+\frac{\kappa_1-\kappa_2}{\kappa_1+\kappa_2}
 A_2\e^{2\eta_2}}
 \biggr].  \no\\
\ee
The potential $u(x,t)$ is written using the superpotential $W(x,t)$ 
as $u(x,t)=W^2(x,t)-\partial_x W(x,t)+E_0$.
The partner potential $\tilde{u}(x,t)=W^2(x,t)+\partial_x W(x,t)+E_0$ 
is calculated to give 
\be
 \tilde{u}(x,t) 
 = \kappa_1^2-\frac{2\kappa_2^2}{\cosh^2(\kappa_2 x-4\kappa_2^3t+\delta_2)},
\ee
where 
\be
 \delta_2 = \frac{1}{2}\ln 
 \left(\frac{\kappa_1-\kappa_2}{\kappa_1+\kappa_2}A_2\right).
\ee
Here, $\tilde{u}$ represents a single-soliton solution.
This means that the double-soliton system is 
a superpartner of the single-soliton system.

Now, we consider counterdiabatic driving for the system
with the double-soliton  potential $u(x,t)$.
The counterdiabatic term is given by 
\be
 \hat{H}_{\rm cd}(t)
  &=& -4\hat{p}^3
 -3\left(\hat{p}u(\hat{x},t)+u(\hat{x},t)\hat{p}\right) \no\\
 &=& -4(\hat{p}+iW(\hat{x},t))\hat{p}(\hat{p}-iW(\hat{x},t)) \no\\
 && -\left(\hat{p}\tilde{u}(\hat{x},t)+\tilde{u}(\hat{x},t)\hat{p}\right)
 -4E_0\hat{p}.
\ee
When we consider the ground state, the first term is neglected
because of the relation (\ref{zero}).
We have a counterdiabatic term that is linear in $\hat{p}$.
The system is controlled by the scalar and vector potentials.
If we want to control the system using the scalar potential,
we can perform the gauge transformation, as we show below.

The adiabatic state with the ground-state energy is obtained by 
solving Eq.~(\ref{gs}) as 
\be
 &&\psi_{\rm ad}^{(0)}(x,t)=C(t)\e^{\kappa_1 x}\no\\
 &&\times\frac{1+\frac{\kappa_1-\kappa_2}{\kappa_1+\kappa_2}A_2\e^{2\eta_2}}
 {1+A_1\e^{2\eta_1}+A_2\e^{2\eta_2}
 +\left(\frac{\kappa_1-\kappa_2}{\kappa_1+\kappa_2}\right)^2A_1A_2
 \e^{2(\eta_1+\eta_2)}}, \no\\
\ee
where $C(t)$ represents normalization including a phase factor.
The Schr\"odinger equation is written as 
\be
 && i\frac{\partial}{\partial t}\psi_{\rm ad}^{(0)}(x,t) \no\\
 &=& \biggl[-\frac{\partial^2}{\partial x^2}+u(x,t)
 +i\left(\frac{\partial}{\partial x}\tilde{u}(x,t)
 +\tilde{u}(x,t)\frac{\partial}{\partial x}\right)
 \no\\ && 
 -4i\kappa_1^2\frac{\partial}{\partial x}
 \biggr]\psi_{\rm ad}^{(0)}(x,t) \no\\
 &=& \biggl[
 \left(-i\frac{\partial}{\partial x}
 -\tilde{u}(x,t)+2\kappa_1^2\right)^2
 \no\\ &&  
 +u(x,t)-\left(\tilde{u}(x,t)-2\kappa_1^2\right)^2
 \biggr]\psi_{\rm ad}^{(0)}(x,t).
\ee
We perform the gauge transformation
\be
 && \tilde{\psi}_{\rm ad}^{(0)}(x,t)=U(x,t)\psi_{\rm ad}^{(0)}(x,t), \no\\
 && U(x,t)=\exp\left[
 -i\int \diff x\,\left(\tilde{u}(x,t)-2\kappa_1^2\right)\right]
\ee 
to write 
\be
 i\frac{\partial}{\partial t}\tilde{\psi}_{\rm ad}^{(0)}(x,t)
 &=& \left[
 -\frac{\partial^2}{\partial x^2}
 +u(x,t)+V_{\rm cd}(t)
 \right]\tilde{\psi}_{\rm ad}^{(0)}(x,t), \no\\
\ee
where 
\be
 V_{\rm cd}(t)
 &=&-\left(\tilde{u}(x,t)-2\kappa_1^2\right)^2
 +\frac{\partial}{\partial t}\int\diff x\,\tilde{u}(x,t) \no\\
 &=&-\left(\tilde{u}(x,t)-2\kappa_1^2\right)^2
 -4\kappa_2^2\tilde{u}(x,t).
\ee
Thus, the unitary-transformed state $\tilde{\psi}_{\rm ad}^{(0)}(x,t)$
is controlled by the counterdiabatic potential $V_{\rm cd}(x,t)$.

\section{Toda equations}

We derive the Toda equations and Lax form in this section.
The Toda system is described by the classical Hamiltonian
\be
 H = \sum_{n}\left(
 \frac{1}{2}p_n^2+\phi(x_n-x_{n-1})
 \right).
\ee
Many particles are aligned in a chain, and the potential function is given by 
\be
 \phi(r)= \e^{-r}+r.
\ee
The classical equations of motion are written as 
\be
 && \frac{\diff x_n}{\diff t} = p_n, \\
 && \frac{\diff p_n}{\diff t} = \e^{-(x_n-x_{n-1})}-\e^{-(x_{n+1}-x_{n})}.
\ee
We define new variables 
\be
 && a_n = \frac{1}{2}\e^{-(x_{n+1}-x_n)/2}, \\
 && b_n = \frac{1}{2}p_n. 
\ee
Then, we obtain the Toda equations:
\be
 && \frac{\diff a_n}{\diff t} = -a_n(b_{n+1}-b_n), \\
 && \frac{\diff b_n}{\diff t} = -2(a_{n}^2-a_{n-1}^2).
\ee

The Toda equations are written in a Lax form.
We have 
\be
 \frac{\diff L}{\diff t} = [M(t),L(t)], 
\ee
where 
\be
 && L(t) = \bmat{ccccccc}
 b_1 & a_1 & 0 & \cdots&  & & 0 \\
 a_1 & b_2 & a_2 & & & &  \\
 0 & a_2 & b_3 & & & &  \\
 \vdots & & & \ddots & & & \vdots \\
  & & & & b_{N-2} & a_{N-2} & 0 \\
  &&&& a_{N-2} & b_{N-1} & a_{N-1} \\
 0 &  &   & \cdots & 0 & a_{N-1} & b_N    
 \emat, \\
 && M(t) = \bmat{ccccccc}
 0 & -a_1 & 0 &  \cdots & & & 0 \\
 a_1 & 0 & -a_2 & & & &  \\
 0 & a_2 & 0 & & & &  \\
 \vdots & & & \ddots & & & \vdots \\
  & & & & 0 & -a_{N-2} & 0 \\
  &&&& a_{N-2} & 0 & -a_{N-1} \\
 0 &  &   & \cdots & 0 & a_{N-1} & 0    
 \emat. \no\\
\ee
Here we used the open boundary condition.

\section{Jordan--Wigner transformation}

We consider the Jordan--Wigner transformation to show that 
the one-dimensional noninteracting fermion 
Hamiltonian in Eq.~(\ref{fermion})
is mapped onto the XY spin Hamiltonian in Eq.~(\ref{hxy}). 

The Jordan--Wigner transformation is defined as 
\be
 && \hat{\sigma}_n^x+i\hat{\sigma}_n^y
 = 2\exp\left(i\pi\sum_{m=1}^{n-1}
 \hat{f}_m^\dag \hat{f}_m^{\phantom{\dag}}\right)\hat{f}_n^\dag, \\
 && \hat{\sigma}_n^z = 2\hat{f}_n^\dag \hat{f}_n^{\phantom{\dag}}-1, 
\ee
where $\{\hat{{\bm \sigma}}_n
=(\hat{\sigma}_n^x,\hat{\sigma}_n^y,\hat{\sigma}_n^z)\}$ 
represent the Pauli matrices, and 
$\hat{f}_n$ and $\hat{f}_n^\dag$ represents the fermion operators.
The fermion operators satisfy anticommutation relations
\be
 && \hat{f}_n^{\phantom{\dag}}\hat{f}_m^\dag
 +\hat{f}_m^\dag\hat{f}_n^{\phantom{\dag}} = \delta_{nm}, \\
 && \hat{f}_n^{\phantom{\dag}}\hat{f}_m^{\phantom{\dag}}
 +\hat{f}_m^{\phantom{\dag}}\hat{f}_n^{\phantom{\dag}}
 = 0.
\ee
These relations are the conditions for the spin operators to satisfy 
the required commutation relations.
Using this transformation, we show that 
Eq.~(\ref{fermion}) is equivalent to Eq.~(\ref{hxy}). 

\section{Spin transport by solitons}

\begin{center}
\begin{figure}[t]
\begin{center}
\includegraphics[width=0.8\columnwidth]{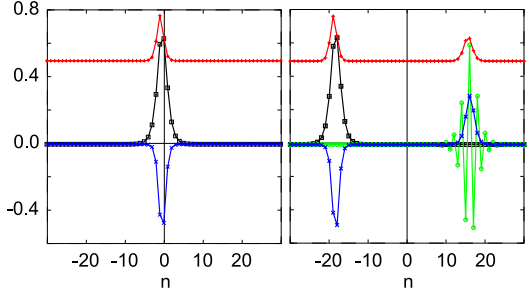}
\end{center}
\caption{(Color online). Soliton solutions to the Toda equations.
for single soliton (left)
and double solitons (right).
Plotted are $J_n(t)$ (red lines with $+$), 
$h_n(t)$ (blue with $\times$), 
the ground-state wavefunction (black with boxes),  
and the first excited-state wavefunction 
(green with circles) at a fixed $t$.}
\label{fig3}
\end{figure}
\end{center}

\begin{center}
\begin{figure}[t]
\begin{center}
\includegraphics[width=0.8\columnwidth]{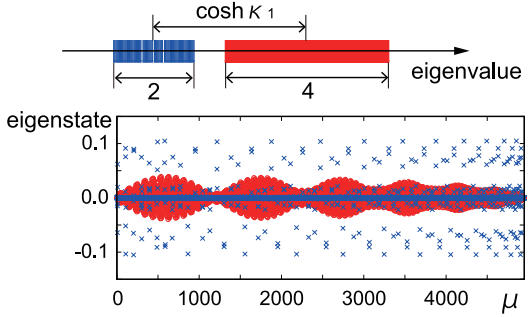}
\end{center}
\caption{(Color online). 
Eigenvalues and eigenfunctions in the double spin-flip sector
for a single soliton.
The continuum states give an energy band of width 4
in the upper panel, and
one of the corresponding eigenfunctions, 
$\psi_\mu$ ($\mu=1,2,\dots,N(N-1)/2$), 
denoted by red circles in the lower panel,  
is extended over the basis states.
The discrete states give a band of width 2, 
and one of the eigenfunctions, 
denoted by blue crosses, is localized in specific states.
We take $\kappa_1=2$ and $N=100$.}
\label{fig4}
\end{figure}
\end{center}

For Toda equations in the infinite system,
the simplest single-soliton solution is given by   
\be
 && J_n(t)=
 \frac{1}{2}\sqrt{\frac{1+\frac{c_1^2(t)z_1^{2n}}
 {1+\frac{c_1^2(t)}{1-z_1^2}z_1^{2(n+1)}}}
 {1+\frac{c_1^2(t)z_1^{2(n+1)}}{1+\frac{c_1^2(t)}{1-z_1^2}z_1^{2(n+2)}}}}, \no\\
 && h_n(t)= \frac{1}{2}\left(
 \frac{c_1^2(t)z_1^{2n+1}}
 {1+\frac{c_1^2(t)}{1-z_1^2}z_1^{2(n+1)}}
 -\frac{c_1^2(t)z_1^{2n-1}}
 {1+\frac{c_1^2(t)}{1-z_1^2}z_1^{2n}}\right), \no\\
\ee
where $z_1=\e^{-\kappa_1}$ and $c_1(t)=c_1(0)\e^{t\sinh\kappa_1}$ 
with positive $\kappa_1$.
$J_n(t)$ and $h_n(t)$
have a single peak around $\kappa_1n\sim t\sinh\kappa_1$.
The wavefunction is localized around the peak of 
the coupling functions.
We plot the single- and double-soliton solutions in Fig.~\ref{fig3}.

\begin{center}
\begin{figure}[t]
\begin{center}
\includegraphics[width=0.8\columnwidth]{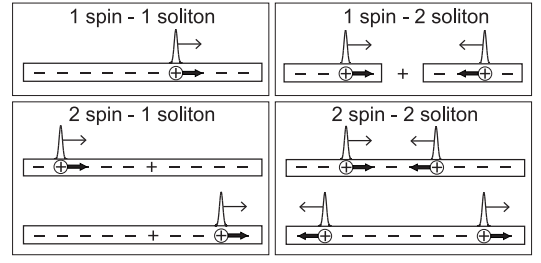}
\end{center}
\caption{Schematic pictures of spin transfer 
for a single soliton (left) and a double soliton (right).
The flipped spins, denoted by the symbol $+$, follow soliton potentials.
The upper (lower) panels show states in the single (double) spin-flip sector.
}
\label{fig5}
\end{figure}
\end{center}

We consider how the eigenstates are controlled 
by the soliton Hamiltonians.
As we discussed in the main body of this letter,
the Hamiltonian commutes with total magnetization 
$\hat{M}=\sum_n\hat{\sigma}_n^z$ and we can take 
the number of flipped spins $M$ as a good quantum number.
In the single spin-flip sectors with $M=\pm(N-2)$, 
there exist continuum ``scattering'' states and discrete ``bound'' states.
The bound-state eigenfunction is trapped in one of the solitons
and is plotted in Fig.~\ref{fig3}.
In the double spin-flip sectors with $M=\pm(N-4)$, 
the eigenvalues and eigenfunctions for a single soliton 
are calculated numerically and are plotted in Fig.~\ref{fig4}.
These results show that the flipped spin is transferred by the solitons
as shown in Fig.~\ref{fig5}.
We find that, to transfer flipped spins efficiently, 
we need not only to apply the magnetic fields $h_n(t)$
but also to change the interactions $J_n(t)$ 
from the uniform value.

\section{Inverse engineering}

We consider inverse engineering of the XY spin Hamiltonian 
\be
 \hat{H}(t)
 = \sum_{n=1}^N\frac{d_n(t)}{2}\left(
 \hat{\sigma}^x_n\hat{\sigma}^x_{n+1}+\hat{\sigma}^y_n\hat{\sigma}^y_{n+1}\right) 
 +\sum_{n=1}^N\frac{h_n(t)}{2}\hat{\sigma}_n^z. 
 \no\\
\ee
Instead of imposing an additional term, 
we construct the invariant $\hat{F}(t)$ so that 
$\hat{H}(t)$ and $\hat{F}(t)$ satisfy Eq.~(\ref{di}).
By choosing $\hat{F}(t)$ properly, 
we design the coupling functions $d_n(t)$ and $h_n(t)$ 
in the Hamiltonian.

The form of the invariant is inferred from the result 
from counterdiabatic driving.
We set 
\be
 && \hat{F}(t) 
 = \sum_{n=1}^N\frac{a_n(t)}{2}\left(
 \hat{\sigma}^x_n\hat{\sigma}^x_{n+1}
 +\hat{\sigma}^y_n\hat{\sigma}^y_{n+1}\right)
 \no\\ &&
 +\sum_{n=1}^N\frac{b_n(t)}{2}\left( 
 \hat{\sigma}_n^x\hat{\sigma}_{n+1}^y
 -\hat{\sigma}_n^y\hat{\sigma}_{n+1}^x\right)  
 +\sum_{n=1}^N\frac{c_n(t)}{2}\hat{\sigma}_n^z, \no\\ \label{Fxy}
\ee 
where $a_n(t)$, $b_n(t)$, and $c_n(t)$ are determined below.
Substituting these expressions into Eq.~(\ref{di}), we obtain 
\be
 && \frac{\diff a_n(t)}{\diff t} = -b_n(t)(h_{n+1}(t)-h_n(t)), \\
 && \frac{\diff b_n(t)}{\diff t} = -d_n(t)(c_{n+1}(t)-c_n(t))
 \no\\ &&
 \qquad\qquad\quad +a_n(t)(h_{n+1}(t)-h_n(t)),  \\
 && \frac{\diff c_n(t)}{\diff t} = 
 -2\left(d_n(t)b_n(t)-d_{n-1}(t) b_{n-1}(t) \right),\\
 && d_n(t) a_{n-1}(t) =d_{n-1}(t) a_n(t),  \\
 && d_n(t) b_{n-1}(t) =d_{n-1}(t) b_n(t).
\ee

It is generally a difficult task to solve these equations.
However, we know that the Toda equations are derived from them.
We write the form  
\be
&& a_n(t)= \frac{1}{2}d_n(t) ,\\
&& b_n(t)=- \frac{1}{2}d_n(t) ,\\
&& c_n(t)= h_n(t).
\ee
Then, the above conditions are reduced to 
\be
 && \frac{\diff d_n(t)}{\diff t}=d_n(t)(h_{n+1}(t)-h_n(t)), \\
 && \frac{\diff h_n(t)}{\diff t}=(d_n^2(t)-d_{n-1}^2(t)).
\ee
Replacing $d_n(t)$ with $\sqrt{2}J_n(t)$, 
we obtain the Toda equations (\ref{Todaeqs}).
$\hat{H}(t)$ and $\hat{F}(t)$ are written as  
\be
 \hat{H}(t)
 &=& \sum_{n=1}^N\frac{J_n(t)}{\sqrt{2}}\left(
 \hat{\sigma}^x_n\hat{\sigma}^x_{n+1}+\hat{\sigma}^y_n\hat{\sigma}^y_{n+1}\right) 
 +\sum_{n=1}^N\frac{h_n(t)}{2}\hat{\sigma}_n^z, \no\\ \\
 \hat{F}(t) &=& \sum_{n=1}^N\frac{J_n(t)}{2\sqrt{2}}\left(
 \hat{\sigma}^x_n\hat{\sigma}^x_{n+1}+\hat{\sigma}^y_n\hat{\sigma}^y_{n+1}\right)
 \no\\
 && -\sum_{n=1}^N\frac{J_n(t)}{2\sqrt{2}}\left(
 \hat{\sigma}_n^x\hat{\sigma}_{n+1}^y-\hat{\sigma}_n^y\hat{\sigma}_{n+1}^x\right)
 +\sum_{n=1}^N \frac{h_n(t)}{2}\hat{\sigma}_n^z. \no\\
\ee
This result shows that 
time evolution of the XY model Hamiltonian $\hat{H}(t)$ gives 
an adiabatic passage defined by 
the instantaneous eigenstates of the invariant $\hat{F}(t)$.

We note that this time evolution is essentially equivalent to 
the counterdiabatic driving that we showed in the main text. 
By using the time-independent gauge transformation 
\be
 \hat{U}=\exp\left(-\frac{i}{2}\sum_{n=1}^N\theta_n\hat{\sigma}^z_n \right), 
\ee
with $\theta_{n+1}-\theta_{n}=-\pi/4$,
we obtain 
\be
 \hat{U}^{\dag}\hat{H}(t)\hat{U}
 &=& \sum_{n=1}^N\frac{J_n(t)}{2}\left(
 \hat{\sigma}^x_n\hat{\sigma}^x_{n+1}
 +\hat{\sigma}^y_n\hat{\sigma}^y_{n+1}\right) 
 \no\\ && 
 +\sum_{n=1}^N\frac{J_n(t)}{2}\left(
 \hat{\sigma}^x_n\hat{\sigma}^y_{n+1}
 -\hat{\sigma}^y_n\hat{\sigma}^x_{n+1}\right) 
 \no\\ && 
 +\sum_{n=1}^N\frac{h_n(t)}{2}\hat{\sigma}_n^z, \\
 \hat{U}^{\dag}\hat{F}(t)\hat{U}
 &=& \sum_{n=1}^N\frac{J_n(t)}{2} \left(
 \hat{\sigma}^x_n\hat{\sigma}^x_{n+1}
 +\hat{\sigma}^y_n\hat{\sigma}^y_{n+1}\right)  
 \no\\ &&
 +\sum_{n=1}^N \frac{h_n(t)}{2}\hat{\sigma}_n^z.
\ee 
The former corresponds to 
the total Hamiltonian $\hat{H}_{\rm ad}(t)+\hat{H}_{\rm cd}(t)$,
and the latter corresponds to $\hat{H}_{\rm ad}(t)$ in Eq.~(\ref{hxy}).

To consider the extensions of the Toda equations, we set  
\be
 && b_n(t)= \alpha(t)a_n(t), \\
 && d_n(t)=\beta(t) a_n(t).
\ee
The conditions for the invariant give 
\be
 && \frac{1}{a_n(t)}\frac{\diff a_n(t)}{\diff t}
 = \frac{1}{2(1+\alpha^2(t))}\frac{c_{n+1}(t)-c_n(t)}{a_n^2(t)-a_{n-1}^2(t)}
 \frac{\diff c_n(t)}{\diff t} 
  \no\\ && \qquad\qquad\qquad\quad
 -\frac{\alpha(t)}{1+\alpha^2(t)}\frac{\diff \alpha(t)}{\diff t}, \\
 && \beta(t)=-\frac{1}{2\alpha(t)(a_n^2(t)-a_{n-1}^2(t))}
 \frac{\diff c_n(t)}{\diff t}, \\
 && h_{n+1}(t)-h_n(t)=-\frac{1}{\alpha(t)a_n(t)}\frac{\diff a_n(t)}{\diff t}.
\ee
The Toda equations can be found by imposing 
a time-independent $\alpha$ and $\beta$.
In inverse engineering, we first find the coupling functions 
$a_n(t)$, $b_n(t)$, and $c_n(t)$ in $\hat{F}(t)$
that satisfy the first equation.
Then, $d_n(t)$ and $h_n(t)$ in $\hat{H}(t)$ 
are obtained from the second and third equations.
The advantage of introducing the time-dependent function $\alpha(t)$ is that 
we can set $\alpha(\tau)=0$ at time $t=\tau$, which means that
the second term of Eq.(\ref{Fxy}) vanishes.
In this case, we can impose the condition 
$[\hat{H}(\tau), \hat{F}(\tau) ]=0$ to obtain 
\be
 && \left.\frac{\diff a_n(t)}{\diff t}\right|_{t=\tau}=0, \\
 && \left.\frac{\diff c_n(t)}{\diff t}\right|_{t=\tau}=0.
\ee
By using $\hat{F}(t)$ that satisfies these conditions,
we can find the eigenstate of the Hamiltonian at $t=\tau$.
In standard inverse engineering, 
we further require an additional condition $[\hat{H}(0),\hat{F}(0)]=0$
at the initial time, $t=0$.
However, considering the final conditions is sufficient 
to show that we can find the eigenstate of the Hamiltonian 
at the final time by following the adiabatic passage 
of the invariant $\hat{F}(t)$.

\section{Nonisospectral Hamiltonian}

We have discussed isospectral Hamiltonians 
in the main body of this letter.
This is obtained by setting $\hat{F}(t)=\hat{H}_{\rm ad}(t)$.
In this section, we discuss a case of nonisospectral systems 
where the invariant is given by 
\be
 \hat{F}(t)=\gamma^2(t)\hat{H}_{\rm ad}(t).
\ee
$\gamma(t)$ represents a time-dependent function.
In this case, the eigenvalues of $\hat{H}_{\rm ad}(t)$ are written as 
\be
 E_n(t)=\frac{\gamma^2(0)}{\gamma^2(t)}E_n(0).
\ee
We show in the following that this case 
does not break the integrability of the systems 
and the results from the scale-invariant driving are reproduced.

First, we study the case where
$\hat{H}_{\rm cd}(t)$ includes terms up to first order in $\hat{p}$.
As in the calculations of the first section in the supplemental material, 
we obtain
\be
 \hat{H}_{\rm cd}(t)=
 \frac{\dot{\gamma}(t)}{2\gamma(t)}\left(\hat{x}\hat{p}+\hat{p}\hat{x}\right)
 +v(t)\hat{p}+\epsilon(t),
\ee
where $v(t)$ and $\epsilon(t)$ are arbitrary functions and
the dot symbol denotes the time derivative.
We note that the second and third terms were derived in Eq.~(\ref{firstp}).
The potential $u$ satisfies the equation 
\be
 \left[\frac{\partial}{\partial t}
 +\left(\frac{\dot{\gamma}(t)}{\gamma(t)}x+v(t)\right)
 \frac{\partial}{\partial x}
 +2\frac{\dot{\gamma}(t)}{\gamma(t)}
 \right]u(x,t)=0.
 \no\\
\ee
Solving this equation, we obtain the scale-invariant form 
\be
 u(x,t)=\frac{1}{\gamma^2(t)}u_0\left(\frac{x-x_0(t)}{\gamma(t)}\right),
\ee
where $u_0$ is an arbitrary function and 
$x_0(t)$ are obtained from $v(t)$ by solving the equation 
\be
 \dot{x}_0(t)-\frac{\dot{\gamma}(t)}{\gamma(t)}x_0(t)= v(t).
\ee
Thus, our formulation gives the result from the scale-invariant systems.

It is also possible to extend the calculation to non-scale-invariant systems.
One of the examples is obtained by extending Eq.~(\ref{cubicp}) to 
\be
 \hat{H}_{\rm cd}(t) 
 &=& \frac{a(t)}{4}\left[
 4\hat{p}^3+3\left(\hat{p}u(\hat{x},t)+u(\hat{x},t)\hat{p}\right)
 \right] \no\\
 && 
 +c_1(t)\hat{p}
 +\frac{\dot{\gamma}(t)}{2\gamma(t)}\left(\hat{x}\hat{p}+\hat{p}\hat{x}\right).
\ee
The last term represents the nonisospectral effect.
By substituting this form to the equation for the invariant, we obtain
the equation for $u$ as 
\be
 \left[\frac{\partial }{\partial t}
 +\left(c_1+\frac{\dot{\gamma}}{\gamma}x\right)\frac{\partial }{\partial x}
 +2\frac{\dot{\gamma}}{\gamma} \right]u 
 = -\frac{a}{4}\left(
 6u\frac{\partial u}{\partial x}
 -\frac{\partial^3 u}{\partial x^3}\right). \no\\
\ee
Then, by putting 
\be
 u(x,t)=\frac{1}{\gamma^2(t)}u_0\left(
 z=\frac{x}{\gamma(t)}, s=\frac{t}{\gamma^3(t)}\right),  \label{kdvgamma}
\ee
we obtain 
\be
 \frac{\partial u_0 }{\partial s}
 &=& -\frac{1}{4}\frac{a}{1-3t\frac{\dot{\gamma}}{\gamma}}\left(
 6u_0\frac{\partial u_0}{\partial z}
 -\frac{\partial^3 u_0}{\partial z^3}\right) \no\\
 && -\frac{\gamma^2 c_1}{1-3t\frac{\dot{\gamma}}{\gamma}}
 \frac{\partial u_0}{\partial z}.
\ee
This has the same form as Eq.~(\ref{kdvc}) and it is a straightforward
task to reduce this form to the standard KdV equation.

By using this extension, we can consider a soliton deformation.
Equation (\ref{kdvgamma}) shows that 
the width, depth, and position of solitons can be changed by 
the time-dependent function $\gamma(t)$.
For example, in the double-soliton case in Eq.~(\ref{double}), 
the parameters $\kappa_1$ and $\kappa_2$ are changed as 
\be
 \kappa_1 \to \frac{\kappa_1}{\gamma(t)}, \\
 \kappa_2 \to \frac{\kappa_2}{\gamma(t)}.
\ee

\end{document}